\documentstyle[12pt]{article} 

\def\Msun{M_{\odot \hskip-5.2pt \bullet}}
\def\kms{km s$^{-1}$} 
\def\r{\hangindent=1pc  \noindent}  
\def\kms{km s$^{-1}$}  \def\deg{$^\circ$} 
\def\vlsr{$V_{\rm LSR}$} 
 
\def\Vrot{V_{\rm rot}}  
\def\ha{H$\alpha$}
\def\nii{[N~{\small II}]}
\def\hii{H~{\small II}}
\def\Ha{H$\alpha$}

\begin{document}

\baselineskip 8mm
\title{Central Kinematics and Rotation Curve of\\
the Sb Galaxy NGC 4527 in  CO, H$\alpha$, and [N~II]  Lines}

\author{Yoshiaki SOFUE,$^{1}$ Akihiko TOMITA,$^{2}$\\
Mareki HONMA,$^{1, 3}$ \& Yoshinori TUTUI$^{1}$\\
1. Institute of Astronomy, Univ. of Tokyo, Mitaka, Tokyo
181-8588 \\
2. Faculty of Education, Wakayama University, Wakayama 640-8510\\
3. National Astronomical Observatory, Mitaka, Tokyo 181-8588} 

\date{}

\maketitle 
\centerline{(Recieved 1998 March 23; accepted 1999 August 26)}

\begin{abstract}
\baselineskip 8mm
We have obtained interferometer observations of the central region of
the Sb galaxy NGC 4527 in the CO ($J=1-0$) line emission using the Nobeyama 
Millimeter Array.
We also obtained optical (\ha\, \nii) spectroscopy using the Okayama 188-cm 
reflector along the major axis.  
The kinematical structure and the distribution of \hii\ regions
show symmetry around the nucleus, 
while the distribution of molecular gas is asymmetric.
The molecular-gas mass shares only 10\% of the dynamical mass
in the central 1 kpc radius region.
Using position-velocity diagrams, we have derived a center-to-disk rotation curve.
It rises steeply in the central region, attaining a maximum of about 250 \kms\
at 400 pc radius,  and then decreases to a minimum at 
2 kpc radius, followed by a disk and  outer flat part.
The rotation curve may provide us with the most similar case 
to that of the Milky Way Galaxy.  
 

\end{abstract}

\baselineskip 8mm
\section{Introduction}
 
CO line emission can be an ideal tool to 
derive rotation curves in the central regions of galaxies
(Sofue 1996, 1997; Sofue et al. 1997, 1998, 1999),
because the CO gas is concentrated  in the inner few kpc region 
 (Sofue et al. 1994). 
Since the CO gas coexists  with \hii\ regions,
\ha-line spectroscopy will also be useful to
derive rotation curves at the central region, if the bulge light can be 
subtracted and the extinction is small enough.
In fact, Rubin et al. (1997) have obtained extensive 
\ha\ spectroscopy for Virgo spirals,
and have shown a steep rise of the rotation curves at the central regions
for many galaxies.
On the other hand, outer rotation and kinematics have been investigated
based on optical and HI line observations
(Rubin et al. 1980, 1982; Bosma 1981;  
Mathewson et al. 1996; Persic, Salucci 1995; Persic et al. 1996).  

In order to derive detailed central rotation curves of galaxies,
we have performed CO and \ha\ line spectroscopy for many galaxies 
using high-resolution CO-line observations as well as in the \ha\ and
\nii\  lines. 
In this paper, we present the result for the Sb galaxy NGC 4527 (table 1),
which is one of the CO-rich galaxies with a high molecular-gas 
concentration near the center (Young et al. 1995)  and having
a weak star-burst activity (Soifer et al. 1987).    

\begin{table*}
\begin{center}
\caption{Observed parameters for NGC 4527.} 
\begin{tabular}{cr}\\
\hline
Type & Sb(c) \\
NED position (1950) & RA= 12h 31m 35.4s \\
 &Dec = 02$^\circ$55$'$43$''$\\
CO kinematical center (1950) & RA=12h 31m 34.9s \\
	& Dec=02\deg55$'$45$''$ \\
Apparent magnitude$^\dagger$ $B_T$ & 11.32 \\
Corrected Apparent magnitude $B_T^{0,i}$ & 10.30  \\
Systemic recession velocity, $V_{\rm sys}$ & 1750 \kms \\
Assumed distance, $D$ &22 Mpc ($H_0=75$ \kms\ Mpc$^{-1}$)\\
 & ($10''=1.05$ kpc)\\
Optical inclination angle $i$ & 69\deg \\
Optical position angle&66\deg \\
CO position angle & 80\deg \\
CO kinematical position angle & 55\deg \\ 
\\
CO peak intensity, $I_{\rm co, peak}$ &  $112$ Jy/beam \kms =
 193 K \kms\ \\
CO luminosity,  $L_{\rm co}$ & $3.65 (\pm 0.54) \times 10^8 $
 K \kms\ pc$^2$ \\ 
Molecular hydrogen mass, $M_{\rm H_2}$ & $5.31(\pm 0.78)\times 10^8
 \Msun$ \\
Molecular gas mass,  $M_{\rm mol.gas}$ & $8.72 (\pm 1.28)\times10^8
 \Msun$ ($X=0.61$) \\
Dynamical mass within 1 kpc (10$''$) radius & $7.5\times 10^9\Msun$\\
Far-IR luminosity$^\ddagger$, $L_{\rm FIR}  (42.5-122.5 \mu {\rm m})$ &
  $7.4 \times 10^{10} L_{\odot}$ \\
CO velocity width, $W_{\rm CO}^{0,i}$ &452 \kms \\
\ha\ velocity width,   $W_{\rm H_\alpha}^{0,i}$ & 451 \kms \\
\hline
\label{table1}
\end{tabular}  
\end{center}

\vskip 2mm
$\dagger$ Sandage, Tamman (1981); ~ $\ddagger$ Soifer et al. (1987). 

\end{table*}
\label{tab1} 

\section{Observations}

\subsection{CO Observations} 
 
High-resolution interferometer observations 
were obtained by using  the  Nobeyama Millimeter 
Array (NMA) in the C configuration on 
1994 December 2 and in the D configuration on 1995 March 16.
Flux and phase calibrations were performed by observing the
nearby radio source 3C 273, which had a flux density of 22 Jy
at the observing frequency.
A 1024-channel FX system (a fast-Fourier-transform
spectro-correlator) was used for spectroscopic data
acquisition  with a total bandwidth of 320 MHz (831 \kms).
The data were then averaged in 16 bins of the original frequency channels,
 resulting in 64 channels with a frequency (velocity) resolution of 5 MHz
(13.0 \kms). 
The cleaned maps were then combined
into a cube of intensity data in the (RA, Dec, \vlsr) space.
Using the data cube, we obtained an intensity map, velocity field
map, and position-velocity diagrams along the major axis.
The synthesized beam was $9''.52\times5''.61$, which yielded a
brightness temperature-to-flux density ratio of 1.725 K/(Jy/beam). 
The observational parameters are summarized in table 1.

\subsection{Optical Spectroscopy} 

CCD optical observations were made using 
a Cassegrain spectrometer equipped on
a 188-cm reflector at the Okayama Astrophysical Observatory.
The slit width was $1.''9$, which is similar to the seeing size,
and was put crossing the nucleus at a position angle of 66\deg. 
The CCD chip had $512 \times 512$ pixels;
we have averaged values in every 2 pixels in spatial direction
to increase the signal-to-noise ratio, resulting in $1.''5$ per pixel,
and the spectral dispersion was 0.767 \AA\
(velocity spacing of 34.9 \kms) per pixel.
Taking into account the seeing size, pixel binding, and the tracking
error during observations, we estimate the spatial resolution to be
about $3''$.
Taking into account the slit width, 
the effective spectral resolution was 1.92 \AA, 
which corresponds to a velocity resolution of 87.1 \kms.
We finally used 160 pixels ($4'$) in the spatial direction, and
270 pixels of wavelength range of 207 \AA\ including the \Ha\
and \nii\ line emissions. 
The exposure time was 1000 sec per one spectrum, and we took
three spectra for the galaxy and combined them.

The continuum emission, such as that due to the bulge, has been subtracted
by applying a background-filtering technique (Sofue and Reich 1979).
We first smoothed the original spectrum S$_0$ in the wavelength 
direction by a convolution function of width
$\Delta x \times \Delta \lambda = 1\times 100$, where
$x$ is the position in pixel along the major axis and $\lambda$
the pixels in the wavelength direction.
The convoluted spectrum S$_1$ was subtracted from the original to 
get S$_2$: ${\rm S}_2={\rm S}_0-{\rm S}_1$.
The negative values in S$_2$ was replaced with 0, and S$_2$ is
subtracted from the original to get ${\rm S}_3={\rm S}_0-{\rm S}_2$.
We smooth S$_3$ by the same convolution function, and obtained
S$_4$. Then, we replaced S$_1$ with S$_4$, and applied again the
above procedure, and repeated this procedure for several times.
We finally obtained a `background (continuum)'-subtracted spectrum, 
${\rm S}_{\rm cont.free}={\rm S}_0 - {\rm S}_1 (i-{\rm th})$. 
The spurious lines from the atmosphere, which are at the
rest wavelengths, are also subtracted from the spectrum.

\section{Results}

\subsection{CO-Intensity Distribution and Molecular Gas Mass}

Figure 1 summarizes the observed results, as superposed on the DSS $B$-band image
of NGC 4527.
Figure 2 shows the obtained distribution of the integrated CO-line intensity.
The molecular-gas distribution is elongated in the direction of 
the position angle of 80\deg, about 14\deg\ shifted from  the major 
axis of the galaxy.
The gas distribution comprises two large clumps, and is highly
asymmetric with respect to the center: the western clump is
much brighter than the eastern one.
The peak intensity of $I_{\rm co, peak}=112$ Jy/beam \kms 
= 193 K \kms\ is obtained at RA=12h31m34.52s, Dec=$02^\circ55'45''.0$.
An arm-like feature is extending toward the east from the eastern 
clump,  coinciding with an optical spiral arm. 
A counter part to this arm appears to be present in the western side. 

--- Fig. 1 ---

--- Fig. 2 ---

The total CO luminosity integrated within $40''\times30''$
area centered on the map center  is estimated to be
$L_{\rm co} = 3.65 (\pm 0.54) \times 10^8 $ K \kms\ pc$^2$ 
for an assumed distance of 
$D=22$ Mpc.
This yields a mass of molecular hydrogen gas of 
$M_{\rm H_2}=C L_{\rm co}=5.31(\pm 0.78)\times 10^8 \Msun$;
the total mass of molecular gas is given by 
$M_{\rm mol.gas}=M_{\rm H_2}/X=8.72 (\pm 1.28)\times10^8\Msun$
for $X=0.61$.
Here, the correction factor $X$ from the hydrogen mass to real
gas mass, 
including He and heavier elements, was taken
from Sofue (1995) for the Milky Way Center, and the conversion
factor at the center of a galaxy, 
$C=0.92 \times 10^{20}$
H$_2$ / K \kms, was taken from Arimoto et al. (1996). 

The dynamical mass $M_{\rm dyn}$ can be estimated from the rotation
curve, as obtained in the next section.
The dynamical mass within 1 kpc radius, where the
major CO features are included, is estimated to be $\sim 7.5\times 10^9\Msun$,
which yields a molecular mass fraction in total (dynamical) mass of about 0.12.
The dynamical mass within the 1.5 kpc radius (15$''$) region, where the CO map
is obtained, is $1.7\times 10^{10}\Msun $, yielding a molecular mass
fraction of only 0.051. 
Therefore, the molecular-to-total mass ratio  in the 
central region of NGC 4527  is not particularly high, but is 
approximately the same as that of the mean value in the whole Milky Way.

\subsection{CO Velocity Field}

The distribution of the intensity-weighted mean velocity of the CO-line
emission is shown in figure 3.
Equal-velocity contours around the nucleus are aligned at a position
angle of $-35^\circ$, showing that the rotating disk has the
major axis at 55\deg. This direction is shifted  by about 
$j=11$\deg\ from that of the optical major axis of the galaxy 
(66\deg) in the
opposite sense to that of the elongation of the intensity map.
If we take into account  the inclination of the galaxy, 
$i=69$\deg,
the tilt of the equal-velocity contours from a pure-circular velocity
contours on the plane of the galaxy, $j_0$, is given by
${\rm tan}~j_0={\rm cos}~i~{\rm tan}~j$;
therefore, 
$j_0=4$\deg.
Hence, non-circular velocities of the disk rotation
is estimated to be $V_{\rm rot} {\rm sin}~j_0 \sim 13 $ \kms,
of the order of the velocity resolution of CO observation. 
So, we may safely assume that the rotation of the CO disk
is approximately circular.

--- Fig. 3 ---

\subsection{CO Position-Velocity Diagram} 

Figure 4 shows the obtained position-velocity (PV) 
diagram of the CO line emission along the major axis.
The major PV feature shows a peculiar bend, which can be interpreted as 
being due to two superposed PV features:
(a) a steep component with a velocity gradient as large as
200 \kms/5$''$= 200 \kms/400 pc across the nucleus, which
indicates a molecular disk or ring of radius 3$''$, or 240 pc.
(b) a higher-velocity, larger-radius ring with a velocity
gradient of 450 \kms/$15''=225$ \kms/7.5$''=225$ \kms/600 pc.
In addition to these major features, we can recognize
(c) a lower-velocity, smaller-gradient PV component crossing the major
PV feature, which may be due to foreground/background spiral arms.
This low-velocity component is also clearly visible in the optical
PV diagrams in the \ha\ and \nii\ lines (figure 5).

--- Fig. 4 ---
 
\subsection{\ha\ and [N~II] Position-Velocity Diagrams}

Figure 5 shows the  \ha\ (lower) and  \nii\ (upper) PV diagrams, where
the continuum light has been subtracted by using the 
background-filtering technique.
The rotation velocity in the \ha\ and \nii\ emission lines 
rises very steeply near the nucleus.
The \ha\ PV diagram shows a steep rise in the central region with
a slope of 200 \kms/$4''$ =200 \kms/400 pc, also indicating a ring-like
distribution of \hii\ regions with a radius of 450 pc.
This feature coincides with the innermost, steep velocity gradient CO 
feature. 
In the right panel of figure 4, we superpose the CO and \ha\ PV diagrams.
It is interesting to note that the \ha\ feature is located inside
the high-velocity CO feature (b). 

The \nii\ PV diagram shows about the same  behavior as the \ha\ line.
However, the innermost \nii\  feature shows a higher-velocity component
with a steeper gradient, which is more coincident with the innermost CO
PV feature. This \nii\ feature attains a maximum velocity of about
250 \kms\ in the central 400 pc region, indicating a very steep rise of
the rotation curve in the inner 400 pc, attaining a maximum velocity as 
high as 250 \kms.
Since the \ha\ line could be superposed by a broad Balmer absorption wing
toward a star-forming a disk in a galaxy, the \nii\ line would be more reliable to
trace the innermost kinematics (see Sofue et al. 1997).
Therefore, we adopt the largest rotation velocities for the innermost
region, e.g. within 500 pc.

--- Fig. 5 ---

\subsection{Line Profiles}

Figure 6 shows line profiles of the CO, \ha, and \nii\ lines. 
Note that the \ha\ and \nii\ PV diagrams are similar to each other,
indicating that the \ha\ line profile is not affected by
the AGN activity. 
the CO profile is more complicated compared to the optical profiles,
which may indicate a more irregular distribution of molecular gas
compared to the star-forming regions, which is already
revealed by the asymmetric CO-intensity distribution.
Both the \ha\ and \nii\ optical profiles are characterized by their 
double horn feature, typical for a rotating disk.
The line widths for CO and \ha\ lines are, however, very similar,
both about 450 \kms\ at the 20\% level of the peak intensity, indicating
that molecular gas and star-forming regions are rotating in a similar
manner.

--- Fig. 6 ---

\subsection{CO Asymmetry vs Kinematical Symmetry}

Although the kinematical property is nearly symmetric as can be seen
in the PV diagrams and velocity field, some 
small non-circular motion of $\sim 13$ \kms\ appears to be 
superposed, suggesting a slightly distorted potential and density waves.
Since the gas mass is much smaller than the dynamical mass,
such density waves would cause a non-linear response 
of the gas associated with strong compression,
resulting in an asymmetry of the molecular gas distribution.
Star formation would occur intensely in such compressed gaseous regions.
In fact, NGC 4527 shows a weak starburst. 
The far-infrared luminosity
is $L_{\rm FIR} = 7.4 \times 10^{10} L_\odot$ (Soifer et al. 1987)
corresponding to a star-formation rate of 10~$M_{\odot}$~yr$^{-1}$
(Sauvage, Thuan 1992), much more intense than that of our Galaxy center. 

\subsection{Rotation Curve}

\def\Sobs{\sigma_{\rm obs}} 
\def\Scor{\sigma_{\rm cor}}
\def\SISM{\sigma_{\rm ISM}}
\def\Vrot{V_{\rm rot}}
\def\Vt{V_{\rm t}} 

We derived a rotation curve by applying an envelope-tracing 
method (Sofue 1996) to the present PV diagrams.
The velocity dispersion of the interstellar gas
($\SISM$) and the velocity resolution of observations 
($\Sobs$) were corrected for as the following.
The correction for the finite-velocity resolution, 
$\Sobs=0.5 {\rm FWHM}$ (full width of half maximum), is given by
$\Scor=\Sobs {\rm exp}[-(d/\Sobs)^2]. $
Here, $d$  is the velocity difference  between the
maximum-intensity velocity near the profile edge, 
and the half-maximum velocity (half width of
the intensity slope at the velocity edge). Equation (1) implies that,
if the original velocity profile of the source is sharp enough,
the observed profile becomes the telescope velocity profile, so that 
the correction is equal to the velocity resolution ($\Scor=\Sobs$). 
On the other hand, if the velocity profile of the source is extended
largely ($d \gg \Sobs$), the correction becomes negligible  
($\Scor \sim 0$) and the half-maximum velocity gives the rotation
velocity.

The terminal velocity $\Vt$ is defined by the velocity at which
the intensity becomes equal to
$I_{\rm t}=0.5 I_{\rm max} $
in the PV diagrams.
Then, the rotation velocity is estimated by
$\Vrot=[\Vt-(\Scor^2 + \SISM^2)^{1/2}]/{\rm sin}~i.$
For the present galaxy, we take  $\SISM \sim 7$ \kms,
$\Sobs=13/2=6.5$ \kms for CO, and $\Sobs=87/2=43.5$ \kms for \ha. 

The obtained rotation curve using the CO, \ha\, and \nii\ PV data
is shown in figure 7.
The innermost curve within 500 pc radius
was obtained by adopting the largest values among the three lines,
because the CO data may miss the highest velocities for its lower
angular resolution than optical.
The curve at radius 500 to 1.5 kpc is an average of the three lines, and the 
outer part is based on the \ha\ line alone.
The CO and \ha\ rotation curves derived by this method
show a maximum of 250 \kms\ in the 500-pc ring. 
The rotation velocity, then, decreases to 170 \kms\ at 2 kpc, 
beyond which, it increases gradually to attain a flat rotation of
about 200 \kms\ at radius 5 to 12 kpc. 
The combined rotation curve from the CO and \ha\ data  
is shown by the thick line in the figure.

--- Fig. 7. ---
 
The high-velocity rotation around the center may indicate 
the existence of a massive core with a scale radius
smaller than several times 100 pc, 
possibly in addition to the bulge component obeying an 
exponential or a de Vaucouleurs law.
The rotation curve of NGC 4527 shows a remarkable similarity to 
the rotation curve of the  Milky Way and many other galaxies.
We stress that such a behavior of nuclear rise of rotation curves
will be universal for massive spiral galaxies, with only a few 
exceptions of dwarf and small-mass galaxies  (Sofue et al. 1997). 

\section{Discussion} 

We have obtained interferometer observations of the
CO ($J=1-0$) line emission using the Nobeyama Millimeter Array,
and \ha\ and \nii\  line spectroscopy using the Okayama 188-cm reflector
along the major axis.
While the CO distribution shows asymmetry and 
is elongated in a direction at a position angle of
80\deg, the kinematical position angle inferred from the velocity
field is 55\deg and approximately coincides with the direction of the optical
major axis, suggesting that the non-circular motion is not dominant.
Besides the kinematical structure, the distribution of \hii\ regions in the
PV diagram show symmetry around the nucleus.
The molecular-gas mass inferred from the CO intensity shares only
10\% of the total (dynamical) mass in the central 1 kpc radius region,
and is only 5\% in 1.5 kpc radius region.

We have derived a nucleus-to-disk rotation curve using the CO, \ha, and
\nii\ PV diagrams along the major axis.
The rotation velocity rises steeply within 400 pc of the nucleus
to about 230-250 \kms, indicating a massive core around the nucleus.
The rotation curve, then, decreases to a minimum at 2 kpc,
 followed by a disk and  outer flat part. 
Such a behavior of the rotation curve mimics that of the Galaxy 
and of many other galaxies.  

 \vskip 2mm

The observations were made during the course of common-use programs at  
the Nobeyama Radio Observatory and the 
Okayama Astrophysical Observatory of the 
National Astronomical Observatories.
The authors thank the staff of the observatories
for their kind help and invaluable suggestions during the observations.

\section*{References}
\small
 
\def\r{\hangindent=1pc  \noindent}  

\r  Arimoto N., Sofue Y., Tsujimoto T. 1996, PASJ 48, 275

\r  Bosma A. 1981, AJ 86, 1825  

\r  Mathewson D.S., Ford V.L. 1996, ApJS 107, 97

\r  Persic M., Salucci P. 1995, ApJS 99, 501

\r Persic M., Salucci P., Stel F.  1996, MNRAS 281, 27

\r Rubin V.C., Ford W.K. Jr, Thonnard N. 1982, ApJ 261, 439 

\r  Rubin V.C., Ford W.K. Jr, Thonnard N. 1980, ApJ 238, 471

\r Rubin V.C., Kenney J.D.P., Young J.S. 1997, AJ 113, 1250 

\r  Sandage A., Tammann G.A. 1981, in {\it A Revised 
Shapley-Ames Catalog of Bright Galaxies}, 
Carnegie Institution of Washigton Pubication 635, Washington D.C.

\r Sauvage M., Thuan T.  1992, ApJ 396, L69

\r  Sofue Y. 1995, PASJ 47, 527 

\r Sofue, Y. 1996, ApJ, 458, 120

\r  Sofue Y. 1997, PASJ 49, 17

\r  Sofue Y., Honma M., Arimoto N. 1994, A\&A 296, 33

\r  Sofue Y., Honma M., Tutui Y., Tomita A. 1997, AJ 114, 2428

\r Sofue Y., Reich W. 1979, A\&A 38, 251 

\r Sofue Y., Tomita A.,  Honma M., Tutui Y. and Takeda Y.
        1998, PASJ 50, 427. 

\r Sofue Y., Tutui Y.,  Honma M., Tomita A., Takamiya T., Koda J., 
	Takeda Y.  1999, ApJ. in press 

\r  Soifer B.T., Sanders D.B., Madore B.F., Neugebauer G., Danielson G.E., 
Elias J. H., Lonsdale C. J., Rice W. L.  1987, ApJ 320, 238

\r  Young J.S., Xie S., Tacconi L.,  Knezek P., Vicuso P., 
Tacconi-Gorman L., Scoville N., Schneider S. 1995, ApJS 98, 219 

\newpage

Figure Captions

Fig. 1. 
{Observed results superposed on a DSS $B$-band image of NGC 4527.
}
  
Fig. 2.
{CO ($J=1-0$) line intensity map as obtained by Nobeyama Millimeter
 Array. The contours are every 10\% of the peak intensity ($=169$ K \kms).  
}

Fig. 3. 
{Intensity-weighted mean velocity field (moment 1) of the CO line 
emission. Contours are drawn at velocity intervals of 50 \kms.
The thick contour near the center is at 1750 \kms, and the
velocity increases from right to left (blue-shifted in W side; 
red-shifted  in E side).  
}

FIg. 4.  {Left panel: CO-line position-velocity diagram along the major axis of
NGC 4527. The velocity is corrected for the inclination.
Middle: The white lines in the middle panel indicate major PV features.
Right panel:  \ha\-line PV diagram is superposed in gray on the CO
PV diagram in contours. The CO contours are
every 10\% of the peak temperature of 0.895 K.
}

Fig. 5.
{\ha\ (lower) and \nii\ (upper) 
position-velocity diagrams along the major axis of NGC 4527,
as obtained with the Okayama 188-cm telescope. 
Continuum bulge emission has been subtracted.
The indicated contour levels are in CCD counts after integrating
all observed frames. The intensity scale is not calibrated because
of the unknown extinction. Note higher velocities in the \nii\ line
for the innermost regions. 
}

Fig.6.
{Line profiles of the CO line (top), \ha and \nii\ lines averaged along
the whole major axis (middle), and the same for the central $80''$ region
(bottom) 
}

Fig. 7.
{Rotation curve of NGC 4527, as obtained from the observed PV diagrams 
in the CO, \ha, and \nii\ lines.
}

\label{last}
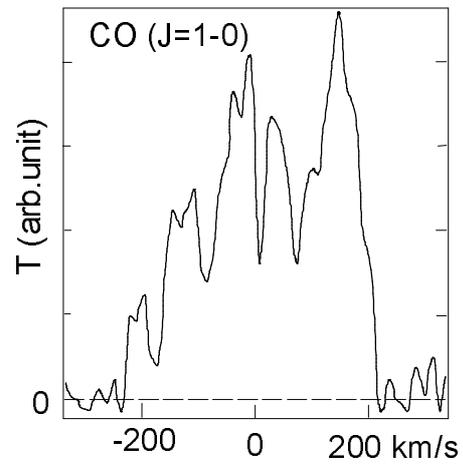
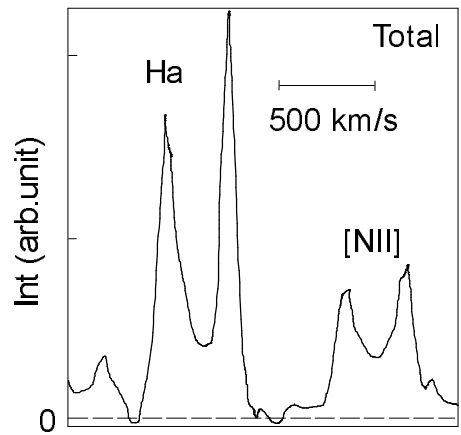
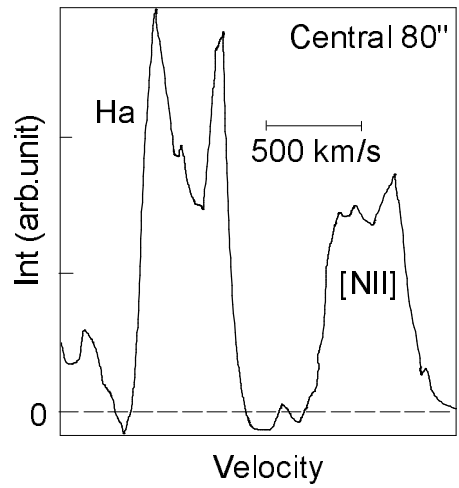
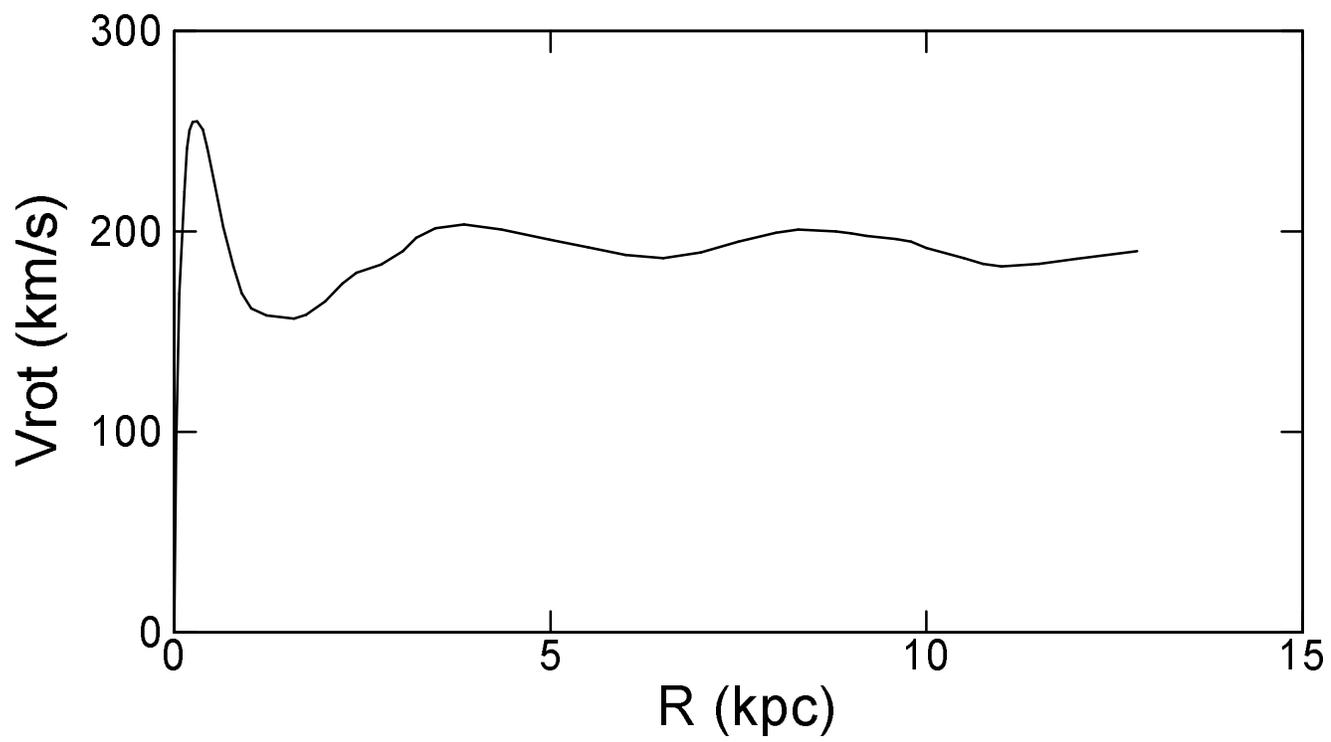
\end{document}